# ERCPMP: An Endoscopic Image and Video Dataset for Colorectal Polyps Morphology and Pathology

Mojgan Forootan[1], Mohsen Rajabnia[2], Ahmad R Mafi[1], Hamed Azhdari Tehrani[3], Erfan Ghadirzadeh[1], Mahziar Setayeshfar[4], Zahra Ghaffari[1], Mohammad Tashakoripour[1], Mohammad Reza Zali[1], Hamidreza Bolhasani[5*]

*Abstract—* In the recent years, artificial intelligence (AI) and its leading subtypes, machine learning (ML) and deep learning (DL) and their applications are spreading very fast in various aspects such as medicine. Today the most important challenge of developing accurate algorithms for medical prediction, detection, diagnosis, treatment and prognosis is data. ERCPMP is an Endoscopic Image and Video Dataset for Recognition of Colorectal Polyps Morphology and Pathology. This dataset contains demographic, morphological and pathological data, endoscopic images and videos of 191 patients with colorectal polyps. Morphological data is included based on the latest international gastroenterology classification references such as Paris, Pit and JNET classification. Pathological data includes the diagnosis of the polyps including Tubular, Villous, Tubulovillous, Hyperplastic, Serrated, Inflammatory and Adenocarcinoma with Dysplasia Grade & Differentiation. The current version of this dataset is published and available on Elsevier Mendeley Dataverse and since it is under development, the latest version is accessible via: https://databiox.com.

*Keywords*- Colorectal Polyps, Dataset, Endoscopy, Colonoscopy, Morphology, Surface Pattern, Pathology, Artificial Intelligence

## I. INTRODUCTION

Colorectal cancer (CRC) is a significant cause of mortality worldwide, responsible for an estimated 1.9 million new cases and 935,000 deaths globally among 5.2 million diagnosed cases in 2020 as shown in Fig. 1 [1]. It is the third most prevalent malignancy worldwide and the second major cause of cancer-related mortality [1]. Detecting CRC early through screening methods like colonoscopy, fecal occult blood tests, and sigmoidoscopy is crucial for improving patient outcomes, which can detect polyps and early-stage malignancies that can be excised before they progress [2, 3].

Colorectal polyps are atypical growths found in the colon or rectum, often discovered during routine colonoscopy exams [4]. Most CRCs develop from precancerous adenomatous polyps [4, 5]. It has been demonstrated that early diagnosis and excision of precancerous colorectal polyps dramatically reduces the risk of colorectal cancer. The excision of such polyps during colonoscopy can prevent the development of cancer from these lesions [3].

Determining if a polyp is cancerous or not relies on the pathological report of histological slides obtained from it; however, certain characteristics observed during colonoscopy can suggest the invasiveness of the polyps [6]. Accurate diagnosis and characterization of these polyps are critical for optimal therapeutic care, as certain endoscopic morphological characteristics of colorectal polyps may point to a greater risk for advanced pathology, such as dysplasia or neoplasia [6, 7]. The neoplastic and benign characteristics of polyps can be viewed and evaluated during a colonoscopy [8]. Neoplastic polyps have the potential to develop into cancer, whereas benign polyps do not [5, 9]. The polyp's size, form, and position might also be crucial in deciding whether it should be removed or monitored [9]. Polyps can be categorized according to their morphological features and based on latest international gastroenterology classification references such as Paris, Pit, and JNET [9].

[1] Gastroenterology and Liver Disease Research Center, Research Institute for Gastroenterology and Liver Diseases, Shahid Beheshti University of Medical Sciences, Tehran, Iran.
[2] Alborz University of Medical Sciences, Alborz, Iran.
[3] Department of Hematology and Medical Oncology, Shahid Beheshti University of Medical Sciences, Tehran, Iran.
[4] Iran University of Medical Sciences
[5] Department of Computer Engineering, Science and Research Branch, Islamic Azad University, Tehran, Iran.
* Corresponding Author: hamidreza.bolhasani@srbiau.ac.ir

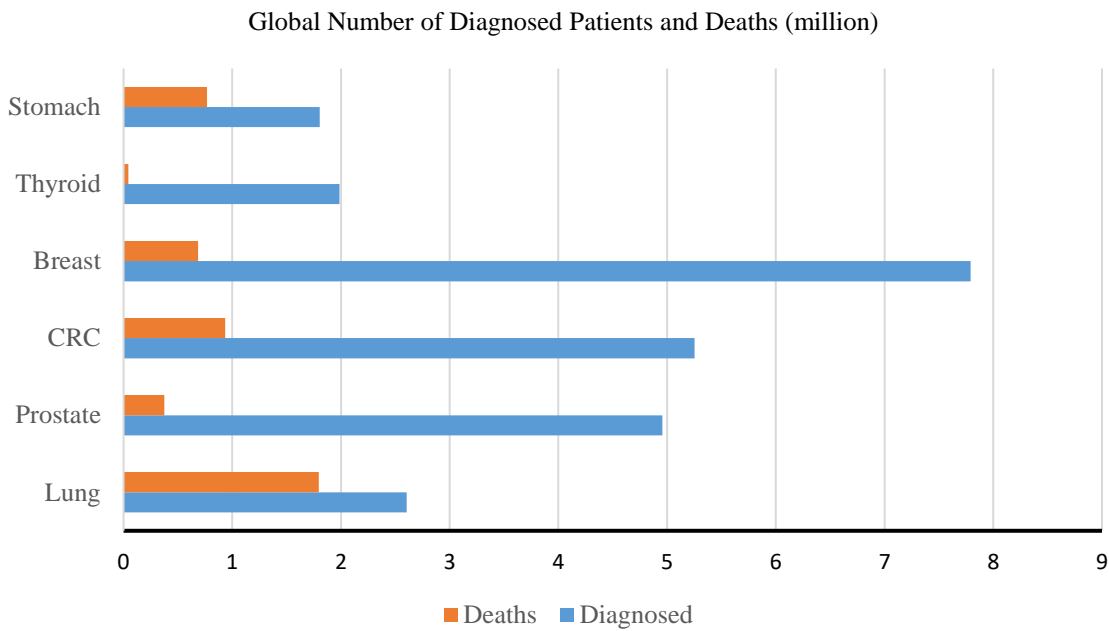

Fig. 1. Diagnosed patients and deaths of five most common cancers based on GLOBOCAN 2020.

In recent times, considerable endeavors have been undertaken to anticipate and identify various forms of cancer by utilizing artificial intelligence (AI) and its subfields, such as machine learning and deep learning [10-12]. The initial crucial phase towards accomplishing this objective involves obtaining an appropriate dataset. Consequently, this study sought to create a meticulously structured collection of images and videos encompassing demographic information, histopathological attributes (including grading, differentiation, and diagnosis), and morphological characteristics (such as size, circumference, Paris class, Pit pattern, JNET classification, and LST type) of colorectal polyps.

## II. RELATED WORKS

The ERCPMP is a unique dataset that offers a valuable resource for researchers and medical professionals. The novelty of this dataset lies in its presentation of both videos and images of colorectal polyps, which allows for a more comprehensive understanding of the polyps' nature. Furthermore, the dataset also presents both morphological and pathological features of the polyps, providing a comprehensive representation of these growths. The combination of these features makes the ERCPMP dataset an essential tool for researchers and medical professionals seeking to better understand colorectal polyps and develop more effective treatments. Additionally, the dataset is well organized, making it easy to use and navigate, further enhancing its value as a research tool. In addition to the aforementioned features, it's worth noting that ERCPMP is unique in that no datasets with high similarity were found in the literature. This further underscores the novelty and importance of your work in providing a comprehensive and well-organized dataset for studying the morphological and pathological features of colorectal polyps. The absence of similar datasets highlights the need for this work and the potential impact it can have on the field of gastroenterology (Fig 1, and 2).

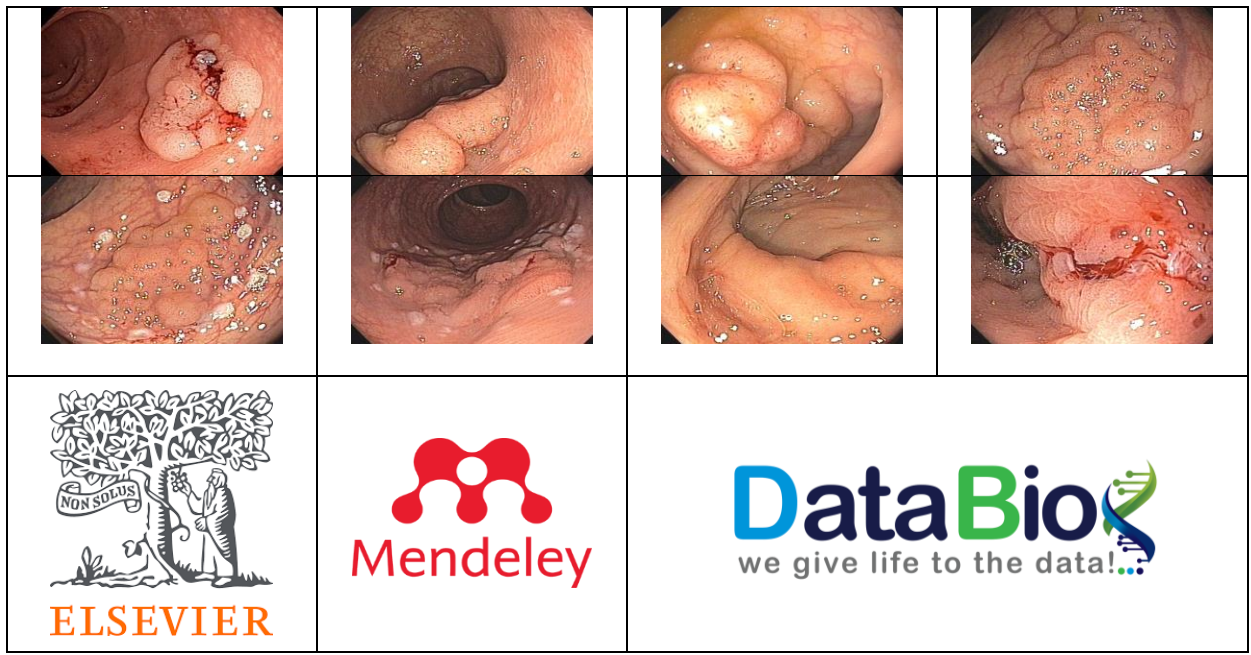

Fig. 1. Some image samples from ERCPMP dataset

Some dataset such as "CVC-ClinicDB", "ACU-Mayo Clinic Colonoscopy Video Database", and "Kvasir-SEG" are publicly available which contain polyp images or videos (or both) [13]. These datasets are similar to ERCPMP in terms of morphological assessment; however, some of them lack labeling of morphologic features and some lack the exclusivity of including histopathology. "CVC-ClinicDB" was organized by Bernal et al. [14] and contains 612 images from 29 various sequences of colonoscopy videos. "ACU-Mayo Clinic Colonoscopy Video Database" was established by Tajbakhsh et al. [15], which is a video dataset containing 20 colonoscopy videos (10 videos with colorectal polyps and 10 without).

Picon et al. [16] represented a dataset named "PICCOLO HE/MPM Image Collection" for colorectal polyps' histopathological classification using hematoxylin-eosin and multi-photon microscopy images. The authors organized a dataset of 50 colorectal polyp samples. Images are presented in several resolutions and classified based on both pathologic classes (healthy, hyperplastic/adenoma, and adenocarcinoma) and dysplasia grade (low or high).

Chattree et al. [17] provided a 23-parameter dataset named "the ACPGBI/BSG Complex Colorectal Polyp Minimum Dataset" which includes demographic details, morphological description and polyp surface features, and histopathologic slides only in a total of 40 patients with large non-pedunculated colorectal polyps.

Borgli et al. [18] developed a general dataset for gastrointestinal (GI) endoscopy named "HyperKvasir". This dataset comprises 10,662 labeled and 99,417 unlabeled images, and 374 videos, and shares many normal and abnormal finding during a GI endoscopy. HyperKvasir is a large and general dataset for GI endoscopy; however, the ERCPMP dataset offers more specific images and videos regarding colorectal polyps morphology and histopathology.

The BioBank foundation released a public dataset named "PICCOLO RGB/NBI (Widefield) Image Collection" which is comprised of 3433 images in the PNG format, including narrow band and wight light, from 76 polyps in 48 different patients. The dataset also provides information of a polyp's size and paris classification, as well as histological diagnosis. The ERCPMP dataset offers more morphological and histological information compared to this dataset [19].

## III. ERCPMP Dataset

ERCPMP is the name of the prepared image and video dataset of this research. This is a histopathological and morphological image and video dataset of 191 patients diagnosed with colorectal polyps including 796 images and 21 videos in total. These numbers are related to the current version, but it is under development to bring more data in the next versions. Images were captured with Olympus colonoscope and are presented in RGB format, JPG type with the resolution of 368 * 256 pixels and 96 dpi. Videos were captured with the same device and are presented in MP4 type. A summary of abbreviations and definitions of terms that were used in this dataset is presented in Table 1 and technical information of the dataset is presented in Table 2. File names in the dataset are based on patient codes provided in the accompanied excel file. The excel file contains information on each patient's demographic data in addition to each polyp morphological and histopathological labeling. Table 3 demonstrates what information was included for each polyp.

| Patient Demographic | | | | Anatomicl Features | | | | Morphology | | Surface Pattern | | Pathology | |
|---|---|---|---|---|---|---|---|---|---|---|---|---|---|
| Patient Code | Image & Video | Sex | Age | Polyp Location | Size (mm) | Circum | Cross Two Folds | Paris | LST Type | Pit | JNET | Diagnosis | Dysplasia Grade & Differentiation |
| 23001 | + | F | 35 | Rectum | 2*2 | <1/3 | Neg | 0-IIa | LST-G HT | III | 2A | Tubular | LGD |
| 23002 | + | M | 76 | Rectum | 3*3 | <1/3 | Pos | 0-IIa | LST-G HT | III, IV | 2A | T + V | LGD |
| 23003 | + | F | 66 | Rectum | 4*3 | <1/3 | Pos | 0-IIa | LST-G HT | IV, V | 2B | Villous | HGD |
| 23004 | - | F | 50 | Rectum | 3.5*1.5 | <1/3 | Pos | 0-IIa + Is | LST-G MN | III, IV, V | 2B | T + V | HGD |
| 23005 | + | F | 73 | Rectum | 4*3 | <1/3 | Pos | 0-IIa + Is | LST-G MN | IV, V | 2B | Villous | HGD |
| 23006 | + | M | 82 | Rectum | 4*3 | <1/3 | Pos | 0-IIa + IIc | LST-NG PD | III, IV | 2A | T + V | LGD |
| 23007 | + | F | 42 | Rectum | 3.5*2 | <1/3 | Pos | 0-IIa + Is | LST-G MN | IV | 2A | Villous | LGD |
| 23008 | + | F | 48 | Rectosigmoid | 4*4 | <1/3 | Pos | 0-IIa | LST-G HT | III, IV | 2A | Serrated | T + V |
| 23009 | Image | M | 68 | Rectosigmoid | 5*3 | >1/3 | Pos | 0-IIa | LST-G HT | III, IV, V | 3 | T + V | HGD |
| 23010 | Image | F | 64 | Rectum | 2*1.5 | <1/3 | Neg | 0-Ips | - | III, V | 2B | Tubular | HGD |
| 23011 | - | F | 53 | Rectum | 1.5*1 | <1/3 | Neg | 0-Is | - | III | 2A | Tubular | LGD |
| 23012 | Image | M | 47 | Rectum | 1.5*1 | <1/3 | Neg | 0-Ips | - | III, IV | 2A | T + V | LGD |
| 23013 | Image | F | 73 | Rectum | 2*2 | <1/3 | Neg | 0-IIa | LST-G HT | III, IV, V | 2B | T + V | HGD |
| 23014 | Image | F | 60 | Rectum | | <1/3 | Neg | - | - | - | - | Adenocarcinoma | N/A |
| 23015 | Image | F | 44 | Rectum | 2.5*2 | <1/3 | Neg | 0-IIa + Is | LST-G HT | III, IV, V | 2B | T + V | HGD |
| 23016 | + | F | 57 | Rectum | 1 | <1/3 | Neg | 0-Is | - | I | 1 | Hyperplastic | - |
| 23017 | Image | F | 43 | Rectum | 1 | <1/3 | Neg | 0-Is | - | I | 1 | Hyperplastic | - |
| 23018 | Image | M | 67 | Rectum | 1*1 | <1/3 | Neg | 0-Is | - | I | 1 | Hyperplastic | - |
| 23019 | + | F | 34 | Rectum | 1*1 | <1/3 | Neg | 0-Is | - | III | 2A | Tubular | LGD |
| 23020 | Image | M | 69 | Rectum | 1.5*1 | <1/3 | Neg | 0-Ip | - | III | 2A | Tubular | LGD |
| 23021 | Image | F | 64 | Rectum | 1*1 | <1/3 | Neg | 0-Is | - | I | 1 | Hyperplastic | - |
| 23022 | - | M | 44 | Rectum | 1*1 | <1/3 | Neg | 0-Is | - | I | 1 | Hyperplastic | - |
| 23023 | Image | F | 32 | Rectum | 3*3 | <1/3 | Pos | 0-IIa | - | III, IV, V | 2B | T + V | HGD |
| 23024 | - | F | 37 | Rectum | 1.5*2 | <1/3 | Neg | 0-IIa | - | - | - | Inflammatory | - |

Fig. 2. Scheme of ERCPMP dataset including patient demographic, anatomical features, morphology and pathology results

The ERCPMP dataset is distinct from similar datasets due to the presence of both morphological and histopathological characteristics of colorectal polyps in addition to including more polyp samples with various features. Generally, six necessary steps were implemented to arrange this dataset as listed:
1. Patients with colorectal polyps were diagnosed and their demographics were recorded
2. Polyp anatomical features, morphology features, and surface pattern were assessed and classified
3. Polyp samples were referred for histopathologic assessment
4. Histologic diagnosis and grading were recorded
5. A written informed consent was obtained from patients to include their clinical details
6. The dataset was organized

Thorough information regarding the quantity of different data in the ERCPMP dataset based on morphological and histopathologic characteristics are presented in Table 4.

## IV. Anatomical, Morphological Features And Surface Pattern

Anatomical features (location, size, circumference, cross two folds), morphological features (Paris classification, LST), and surface pattern (Kudo pit pattern, JNET) play a pivotal role in determining the clinical management and prognosis of colorectal polyps and includes many different classifications systems. Morphological features and surface pattern (Paris classification, LST, and JNET) were the basis for selecting the lesion resection method (Endoscopic Polypectomy, EMR, ESD, FTRD).

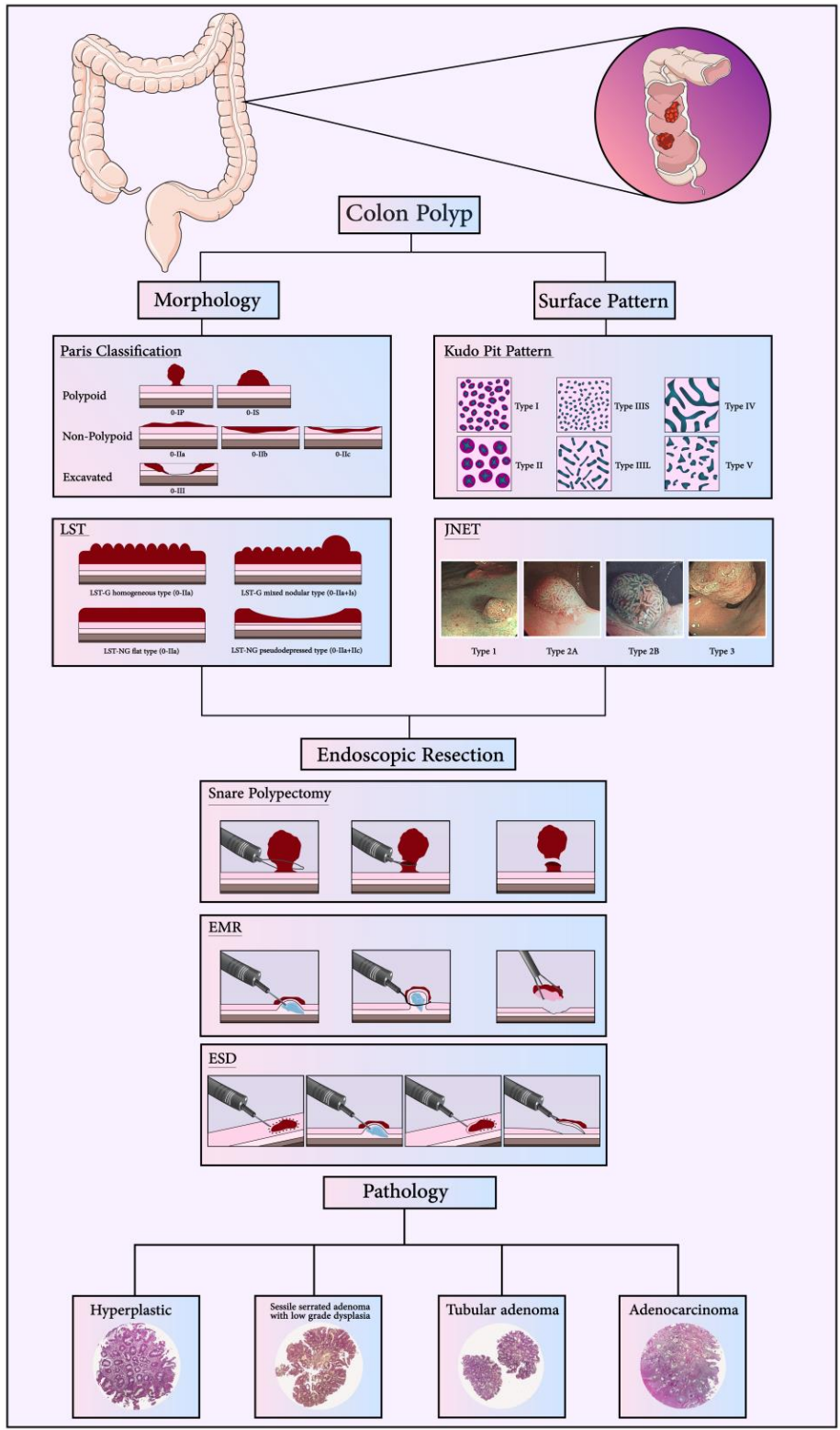

Fig. 3. Comprehensive algorithm used in preparing ERCPMP

- **Paris Classification**

The Paris classification is an endoscopic classification system that focuses on the macroscopic appearance of colorectal polyps. It categorizes polyps into four types based on their morphology: polypoid (0-I and 0-IIa), non-polypoid (0-IIb and 0-IIc), and depressed (0-III and 0-Is) (Fig 3). This classification system is a standardized terminology that aids various medical specialists to communicate with each other, and assists endoscopists in determining the appropriate resection technique and predicting the risk of submucosal invasion [20, 21].

- **LST**

The LST classification system recognizes two major categories of LSTs which are lesions that grow laterally instead of upward or downward [20, 26]:
  i. LST-G:
  - LST-G
  - LST-G-H (Paris 0-IIa)
  - LST-G-NM (Paris 0-IIa and 0-Is)
  ii. LST-NG:
  - LST-NG
  - LST-NG-FE (Paris 0-IIa)
  - LST-NG-PD (Paris 0-IIa and 0-IIc)

- **Pit pattern**

Pit pattern refers to the microscopic appearance of the glandular structures within a colorectal polyp observed during magnifying endoscopy. It helps to distinguish between different histological types of polyps, aiding in risk stratification and treatment decision-making. There are several pit pattern classification systems; however, the Kudo classification is one of the most widely used. It categorizes pit patterns into five main types (from I to V) (Fig 3) [20, 24, 25].

- **JNET**

JNET classification system is a comprehensive and widely used endoscopic classification specifically designed for colorectal polyps. It incorporates the use of magnifying endoscopy with narrow-band imaging and provides a standardized framework for assessing the morphological characteristics of polyps. The classification is based on the evaluation of surface pattern, color, vessel pattern, and microsurface structure [22, 23].

The JNET classification includes three surface pattern categories; type 1 (homogenously colored), type 2A (heterogeneously colored), and type 2B (excavated appearance). The JNET classification incorporates three color categories; light blue which indicates the presence of a superficial layer of epithelial cells, light blue-white which suggests the absence of the superficial layer and exposure of the lamina propria, and dark blue which indicates the presence of submucosal vessels. Also, the JNET classification system identifies four vessel pattern categories; type 1 (well-organized and regularly distributed vessels), type 2A (slightly distorted), type 2B (clearly distorted), type 3 (absence of any visible vessels); and two microsurface structures including open (visible crypts openings) and closed (absence of visible crypt openings). The combination of surface pattern, color, vessel pattern, and microsurface structure in JNET system provides valuable information for predicting histological characteristics of polyps and the risk of submucosal invasion, aiding in determining the need for resection or surveillance (Fig 3) [20, 22, 23].

## V. HISTOPATHOLOGIC FEATURES

Pathological evaluation plays a crucial role in the diagnosis, differentiation, and grading of colorectal polyps. Accurate classification and characterization of polyps with a proper pathological assessment provides essential information for patient management and includes both macroscopic and microscopic evaluations. Macroscopic evaluation is the initial step in the pathological examination of colorectal polyps. It involves the assessment of polyp size, shape, color, surface characteristics (smooth or granular), and presence of features such as ulceration or depression. Microscopic evaluation involves the examination of histological features of colorectal polyps under a microscope and includes mainly of the two following aspects; histology, and differentiation and grading [20, 27-29].

**Histological classification**

1. Hyperplastic Polyps: Characterized by normal glandular architecture with minimal cytological abnormalities. These polyps typically show serrated epithelium and elongated crypts.

2. Adenomatous Polyps: Represented by dysplastic glandular structures, which are further classified into:

    - Tubular Adenoma: Predominantly composed of tubular structures with uniform glandular features.

    - Villous Adenoma: Composed of long, slender finger-like projections (villi).

    - Tubulovillous Adenoma: Exhibit a combination of tubular and villous architecture.

**Differentiation and Grading**

1. Dysplasia: Refers to the presence of abnormal cellular changes within the epithelial layer. Dysplasia is classified as:

- Low-grade dysplasia: Mildly disordered glandular architecture with minimal cytological abnormalities.

- High-grade dysplasia: Marked disordered glandular architecture with prominent cytological abnormalities.

2. Carcinoma: Carcinoma in Situ (CIS): Malignant cells confined to the epithelial layer without invasion into the underlying tissue.

- Invasive Carcinoma: Malignant cells invading beyond the epithelial layer into the submucosa or deeper layers.

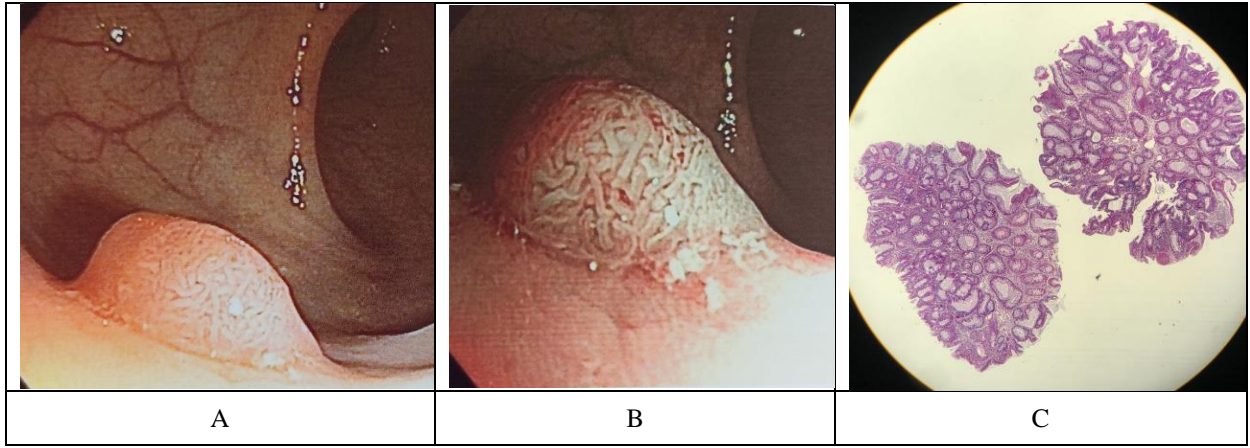

A. White-light colonoscopy showing a sessile (Paris classification Is) polyp with Kudo pit pattern IIIL.
B. Magnified NBI pattern showing JNET classification 2A.
C. Pathology photo with ×20 magnification showing tubular adenoma.

| Anatomical Features | | | | Morphologic Classification | | Surface Classification | | Pathology |
|---|---|---|---|---|---|---|---|---|
| Location | Size | Circumference | Cross Two Folds | Paris Classification | LST | JNET Classification | Kudo Pit Pattern Classification | |
| Rectum | 12mm | <1/3 | Negative | Is | No | 2A | IIIL | Tubular Adenoma |

Fig. 4. A sample of anatomical features, morphologic, surface classification and pathology described in the dataset

## VI. Value of The Data

- ERCPMP is a comprehensive dataset with high quality images and videos of colorectal polyps.

- ERCPMP dataset images and videos are labeled according to both morphological and histopathological features.

- ERCPMP dataset can be utilized as a training dataset in the development of artificial intelligence algorithms.

- ERCPMP dataset could be a valuable source for investigational and educational purposes of medical health professionals.

## VII. Discussion

Recently, significant efforts have been devoted to utilizing AI and its subsets such as machine learning and deep learning to predict and identify various types of neoplasms, with a particular focus on colorectal cancer. Obtaining a well-structured dataset plays a vital role in training machine learning algorithms and is among the initial steps in this matter. Histologic grading and morphological classifications are among the important factors for prognosis and determining the response to treatments like chemotherapy, radiotherapy, and surgery. Examining several notable studies conducted between 2010 and 2023 revealed the absence of a comprehensive histopathological and morphological image and video dataset specifically for colorectal polyps [13-19]. This research gap prompted the development of the ERCPMP dataset. The current version of this dataset is available on Elsevier Mendeley dataset portal at https://data.mendeley.com/datasets/7grhw5tv7n/2 and since it is under development, the latest version is available via: https://databiox.com. The current version of ERCPMP comprises a collection of 796 RGB colonoscopy images in JPG format and 21 colonoscopy videos in MP4. These images and videos were captured from 191 patients diagnosed with colorectal polyps at the Gastroenterology and Hepatology Research Center of Shahid Beheshti University of Medical Sciences in Iran between 2014 and 2019. Each image in the dataset is annotated with its corresponding diagnosed grade, enabling the training of machine learning algorithms for polyp detection.

**Table 1.** Abbreviations and Definitions.

| Abbreviation | Definition | Abbreviation | Definition |
|---|---|---|---|
| **CRC** | Colorectal Carcinoma | **LST-G-H** | Laterally Spreading Tumor, Granular-Homogenous |
| **EMR** | Endoscopic Mucosal Resection | **LST-G-NM** | Laterally Spreading Tumor, Granular-Nodular Mixed |
| **ESD** | Endoscopic Submucosal Dissection | **LST-NG** | Laterally Spreading Tumor, Non-Granular |
| **FTRD** | Full-Thickness Resection Device | **LST-NG-FE** | Laterally Spreading Tumor, Non-Granular-Flat Elevated |
| **JNET** | Japanese Narrow Band Imaging Expert Team | **LST-NG-PD** | Laterally Spreading Tumor, Non-Granular-Pseudo Depressed |
| **LST** | Laterally Spreading Tumor | **NBI** | Narrow-Band Imaging |
| **LST-G** | Laterally Spreading Tumor, Granular | | |

**Table 2.** ERCPMP dataset technical specifications.

|  | Value |  | Value |
|---|---|---|---|
| **Number of Images** | 796 | **Number of Videos** | 21 |
| **Image Format** | RGB | **Video Frame Rate** | 30 fps |
| **Image File Type** | JPG | **Video File Type** | MP4 |
| **Image Resolution** | 96 dpi | **Video Resolution** | 480 * 288 |
| **Size (Pixels)** | 368 * 256 |  |  |

**Table 4.** Data quantity in the ERCPMP dataset for each characteristic.

|  |  | Numbers |  |  | Numbers |  |  | Numbers |
|---|---|---|---|---|---|---|---|---|
| **Patients** |  | 191 | **Circumference** | > 1/3 | 10 | **JNET** | 1 | 34 |
| **Gender** | Male | 71 |  | <= 1/3 | 181 |  | 2A | 87 |
|  | Female | 120 | **Cross Two Folds** | Positive | 22 |  | 2B | 20 |
| **Location** | Ileocecal valve | 2 |  | Negative | 169 |  | 3 | 7 |
|  | Cecum | 4 | **Paris Class** | 0-Is | 70 | **LST-Type** | GMN | 7 |
|  | Ascending | 55 |  | 0-Ip | 10 |  | GHT | 12 |
|  | Hepatic flexure | 4 |  | 0-Ips | 38 |  | NGFT | 7 |
|  | Transverse | 15 |  | 0-IIa | 30 |  | NGPD | 3 |
|  | Splenic Flexure | 13 |  | 0-IIb | 3 | **Diagnosis** | Tubular | 68 |
|  | Descending | 23 |  | 0-IIc | 0 |  | Villous | 6 |
|  |  |  |  | Mix | 0 |  |  |  |
|  | Sigmoid | 39 | **Pit Pattern** | I | 32 |  | Tubulovillous | 31 |
|  | Rectosigmoid | 7 |  | II | 2 |  | Serrated | 4 |
|  | Rectum | 28 |  | III | 63 |  | Hyperplastic | 31 |
|  | Other | - |  | IV | 5 | **Grading** | Low | 81 |
|  |  |  |  | V | 3 |  | Moderate | - |
|  |  |  |  | Mix | 41 |  | High | 23 |

## VIII. DECLARATIONS

### AVAILABILITY OF DATA AND MATERIALS
Available.

### FUNDING
Not Applicable.

### ACKNOWLEDGEMENTS
Not Applicable.

### CONFLICT OF INTEREST

- All authors have participated in (a) conception and design, or analysis and interpretation of the data; (b) drafting the article or revising it critically for important intellectual content; and (c) approval of the final version.

- This manuscript has not been submitted to, nor is under review at, another journal or other publishing venue.

- The authors have no affiliation with any organization with a direct or indirect financial interest in the subject matter discussed in the manuscript

- The following authors have affiliations with organizations with direct or indirect financial interest in the subject matter discussed in the manuscript:

**Authors Biography**

| | |
|---|---|
| **Mojgan Forootan, MD**<br><br>Professor of Gastroentrology<br><br>Shahid Beheshti University of Medical Sciences<br><br>Gastroenterology and Liver Disease Research Center, Research Institute for Gastroenterology and Liver Diseases, Shahid Beheshti University of Medical Sciences, Tehran, Iran | 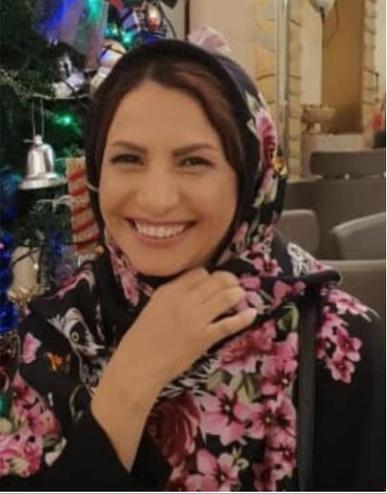 |
| **Mohammad Reza Zali, MD**<br><br>Distinguished Professor of Gastroenterology<br><br>Head of Gastroenterology and Liver Disease Research Center, Research Institute for Gastroenterology and Liver Diseases, Shahid Beheshti University of Medical Sciences, Tehran, Iran | 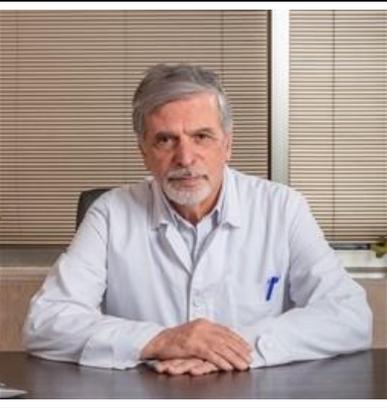 |
| **Hamidreza Bolhasani, PhD**<br><br>AI/ML Researcher / Visiting Professor<br><br>Founder and Chief Data Scientist at DataBiox<br><br>Ph.D. Computer Engineering from Science and Research Branch, Islamic Azad University, Tehran, Iran. 2018-2023.<br><br>Fields of Interest: Machine Learning, Deep Learning, Neural Networks, Computer Architecture, Bioinformatics | 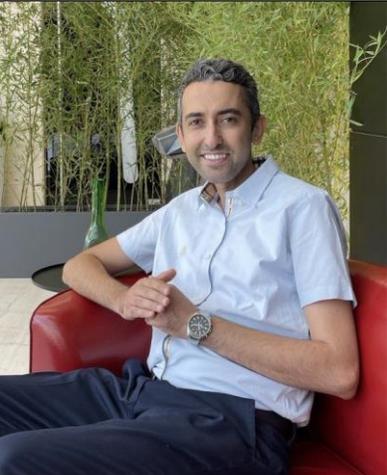 |

**Mohsen Rajabnia, MD**

Assistant Professor of Gastroentrology

Alborz University of Medical Sciences

Gastroenterology and Liver Disease Research Center, Research Institute for Gastroenterology and Liver Diseases, Shahid Beheshti University of Medical Sciences, Tehran, Iran

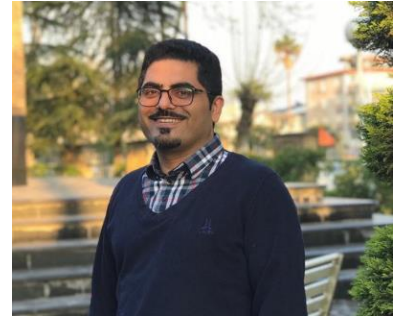

**Ahmad R. Mafi, MD**

Assistant Professor of Radiation Oncology

Shahid Beheshti University of Medical Sciences

Gastroenterology and Liver Disease Research Center, Research Institute for Gastroenterology and Liver Diseases, Shahid Beheshti University of Medical Sciences, Tehran, Iran

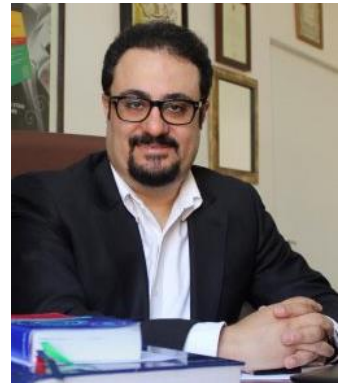

**Hamed Azhdari Tehrani, MD**

Hematologist – Medical Oncologist

Department of Hematology and Medical Oncology,

Shahid Beheshti University of Medical Sciences,

Tehran, Iran

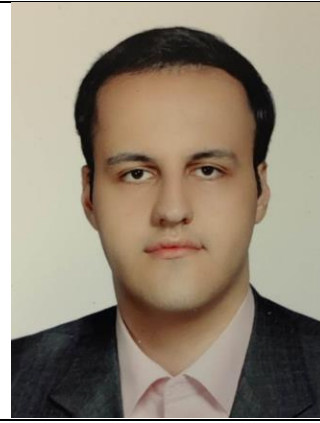

**Erfan Ghadirzadeh, MD**

Mazandaran University of Medical Sciences

Gastroenterology and Liver Disease Research Center, Research Institute for Gastroenterology and Liver Diseases, Shahid Beheshti University of Medical Sciences, Tehran, Iran

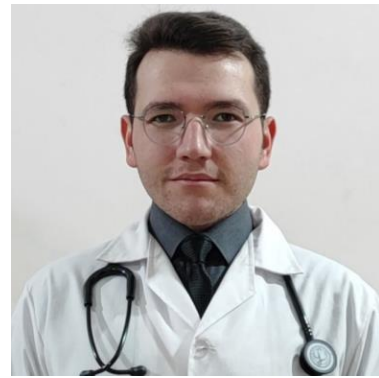

**Mahziar Setayeshfar, Medical Student**

Iran University of Medical Sciences

Gastroenterology and Liver Disease Research Center, Research Institute for Gastroenterology and Liver Diseases, Shahid Beheshti University of Medical Sciences, Tehran, Iran

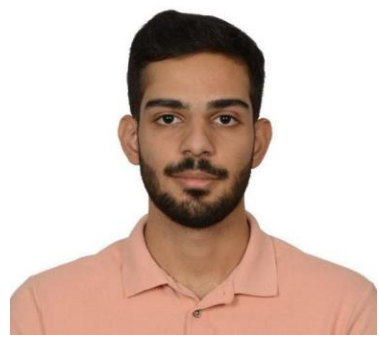

**Zahra Ghaffari, Medical Student**

Shahid Beheshti University of Medical Sciences

Gastroenterology and Liver Disease Research Center, Research Institute for Gastroenterology and Liver Diseases, Shahid Beheshti University of Medical Sciences, Tehran, Iran

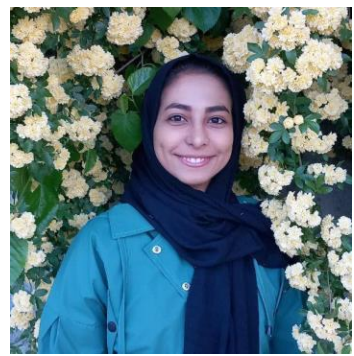

**Mohammad Tashakoripour**

Researcher, Tehran University of Medical Sciences

Gastroenterology Department, Amiralam Hospital, Tehran University of Medical Sciences, Tehran, Iran

Gastroenterology and Liver Disease Research Center, Research Institute for Gastroenterology and Liver Diseases, Shahid Beheshti University of Medical Sciences, Tehran, Iran

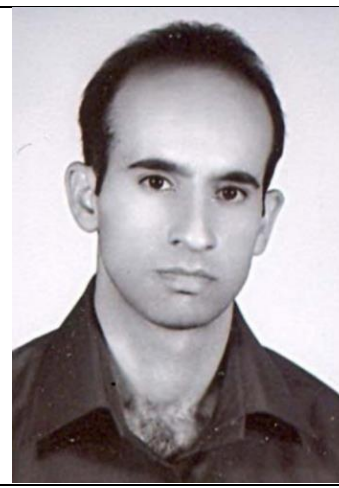